\documentclass{pasj00}
\def\gsim{\ \raise 3pt \hbox{$>$} \kern -8.5pt \raise -2pt \hbox{$\sim$}\ }

\begin{document}
\SetRunningHead{Fleishman}{Pitch-Angle Anisotropy}
\Received{2004/12/31}
\Accepted{2005/01/01}

\title{Radio Emission from Anisotropic Electron Distributions}

\author{Gregory D. \textsc{Fleishman}}%
\affil{National Radio Astronomy Observatory, Charlottesville, VA
22903 USA} 


%


\maketitle

\begin{abstract}
The effect of pitch-angle anisotropy of fast electrons on
generation of nonthermal radio emission is studied. Incoherent
gyrosynchrotron radiation is shown to depend strongly on the
anisotropy. In particular, the spectral index of gyrosynchrotron
radiation increases up to a factor of 3-4 compared with the
isotropic case. The degree of polarization (X-mode) increases for
loss-cone distributions, while decreases for beam-like
distributions. Ample evidence of the pitch-angle anisotropy effect
on (1) spatial distribution of the radio brightness, (2) spatially
resolved light curves of the intensity and polarization, (3)
spectral hardness of microwave bursts, is found by exploring
observations performed with Nobeyama Radioheliograph and Owens
Valley Solar Array. Above some threshold in the angular gradient,
the electron cyclotron maser instability (coherent gyrosynchrotron
emission) can develop, provided that standard gyrosynchrotron
emission is accompanied by a lower-frequency intense coherent
emission. The coherent emission is studied in detail for a
realistic distribution of fast electrons over the energy and
pitch-angle. In agreement with observations of narrowband radio
spikes, it is found that (1) hard (power-law) distributions over
energy are preferable to produce the coherent emission, (2) the
threshold of the instability corresponds to quite an anisotropic
electron distribution, thus, the pitch-angle anisotropy derived
from the properties of the continuum gyrosynchrotron radiation
will not necessarily give rise to coherent emission (either
enhanced isotropization implied by quasilnear saturation of the
instability).
\end{abstract}

\section{Introduction}

Observations of solar radio bursts with Nobeyama Radioheliograph
(NoRH) revealed unambiguously the presence of anisotropic
pitch-angle distributions of fast electrons accelerated in flares
and accumulated in the radio sources. In particular,  the
existence of the microwave loop-top brightness peak of optically
thin gyrosynchrotron (GS) emission \citep{Melnikovetal2001,
Kundu2001} has been interpreted as tracer of a strong
concentration of mildly relativistic electrons trapped and
accumulated in the top of a flaring loop
\citep{Melnikovetal2002a}, which is an evidence of the transverse
anisotropy of radiating electrons. Then, \citet{LeeGary2000}
studied the evolution of the radio spectrum of 1993 Jun 3 flare
observed with Owens Valley Solar Array (OVSA) and found important
indication of the anisotropic injection of fast electrons into the
radio source and their consequent isotropisation due to angular
scattering. The importance of the pitch-angle anisotropy for the
modern radio data interpretation has been widely discussed during
this Nobeyama Symposium
\citep{Altyntsev_2005,Bastian_2005,Gary_2005,Melnikov_2005}. This
paper discusses the effect of the pitch-angle anisotropy on
generation of GS radiation (both incoherent and coherent).

A detailed study (stimulated by the mentioned findings) of
incoherent GS emission produced by anisotropic pitch-angle
distributions of the loss-cone and beam-like types is given in
recent papers \citep{FlMel_2003a,FlMel_2003b}. Accordingly, we
outline only briefly  why the anisotropy has actually a strong
effect on GS emission.

The effect of anisotropy on the optically thin emission is
primarily related to the energy-dependent directivity of the GS
radiation. Indeed, most of the synchrotron radiation by a single
ultrarelativistic electron is emitted within a narrow cone
\begin{equation}
\label{directi} \vartheta \sim \gamma_e^{-1}= mc^2/E
\end{equation}
along the particle velocity, where $\gamma_e$ is the
Lorentz-factor of the electron, $m,~E$ are the mass and energy of
the electron, $c$ is the speed of light, so $\vartheta \ll 1$ if
$\gamma_e \gg 1$. On the other hand, for semirelativistic and
mildly relativistic electrons responsible for the GS emission, the
directivity is not quite as strong, so  the anisotropy (if
present) can significantly change the radiation spectrum when the
transition from moderately relativistic to highly relativistic
regime occurs. Typical energies of fast electrons producing solar
microwave continuum bursts (which are commonly believed to be
produced by GS mechanism, e.g., \cite{BBG}) range from hundreds
keV to a few MeV. Thus, the pitch-angle anisotropy of fast
electrons in solar flares is particularly important for generation
of microwave continuum radio emission, as well as for driving
instabilities, which can result in coherent radio emissions.

\section{Angular Distribution Function}

As a first approximation we consider a (simplified) factorized
distribution function of fast electrons
\begin{figure}
  \begin{center}
    \FigureFile(61mm,62mm){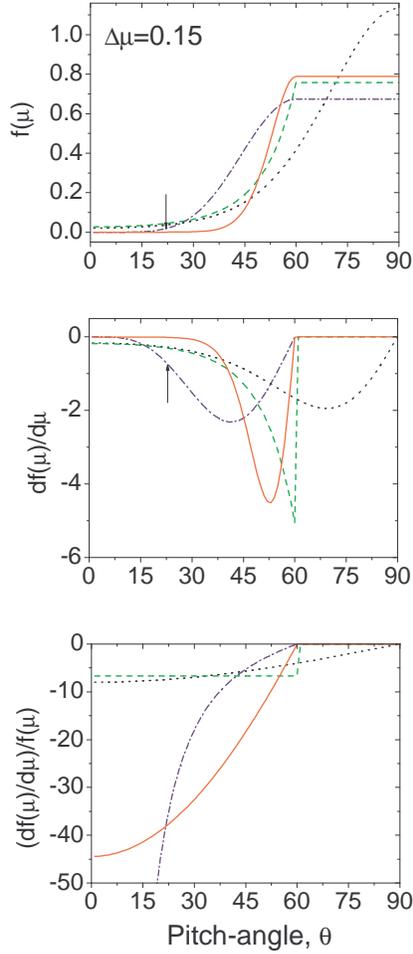}
  \end{center}
  \caption{Top: Angular distribution functions: gaussian \citep{FlYas1994a},
dotted line; sin-N loss-cone \citep{Aschw1990} for $N=6$,
dash-dotted line; exponential loss-cone (\ref{distr_mu_exp}) for
$\Delta \mu = 0.15$, dashed line; and gaussian loss-cone
(\ref{distr_mu_gauss}) for $\Delta \mu = 0.15$, solid line. The
distributions are plotted for $\mu_0=\cos\theta_c=60^o$, all of
them are normalized to unity. Middle: The derivatives of the
angular distributions over $\mu$. Bottom: the ratios of the
derivatives to the distribution functions.}\label{distrib}
\end{figure}

\begin{equation}
\label{distr} f({\bf p}) = {N_e \over 2 \pi} f_1(p) f_2(\mu).
\end{equation}
The distribution function (\ref{distr}) is normalized by $d^3p$
\begin{equation}
\label{norm_1} \int f({\bf p}) p^2 dp d \mu d \varphi = N_e,
\end{equation}
where $N_e$ is the number density of fast electrons, $\mu=cos
\theta$, $\theta$ is the pitch-angle, $\varphi$ is the azimuth
angle, the specific distributions $f_1(p)$ and $f_2(p)$ are
normalized as follows:
\begin{equation}
\label{norm2} \int f_1(p) p^2 dp = 1, \qquad  \int_{-1}^1 f_2(
\mu) d \mu  = 1.
\end{equation}
We assumed the distribution function over the momentum modulus to
be a kind of power-law, while we consider a few different forms
for the pitch-angle distribution to study the sensitivity of the
GS properties to the shape of the angular distribution.

In addition to the gaussian and sin-N functions (explored by
\cite{FlMel_2003b}), we use also the exponential loss-cone
function
\begin{equation}
\label{distr_mu_exp} f_2(\mu) \propto \Biggl\{ \begin{array}{cc}
\exp\left(- {\mu-\cos \theta_c \over \Delta \mu} \right),&
\mu > \cos \theta_c, \\
1, & -\cos \theta_c < \mu < \cos \theta_c\\
\exp\left( {\mu +\cos \theta_c \over \Delta \mu} \right),& \mu <
-\cos \theta_c,
\end{array}
\end{equation}
and the gaussian loss-cone function
\begin{equation}
\label{distr_mu_gauss} f_2(\mu) \propto \Biggl\{ \begin{array}{cc}
\exp\left(- \left({\mu-\cos \theta_c \over \Delta \mu}\right)^2
\right),&
\mu > \cos \theta_c, \\
1, & -\cos \theta_c < \mu < \cos \theta_c\\
\exp\left(- \left({\mu +\cos \theta_c \over \Delta \mu}\right)^2
\right),& \mu < -\cos \theta_c.
\end{array}
\end{equation}

Let us consider these angular distributions (presented in figure
\ref{distrib}) in more detail, having in mind that along with
modification of incoherent GS emission the anisotropy can  give
rise to some instabilities (which depends on relative value of the
angular gradient), in particular, to electron cyclotron maser
(ECM) emission at the low gyroharmonics. The gaussian angular
distribution   provides rapid decrease of the number of electrons
toward the direction of the magnetic field, and affect the GS
substantially \citep{FlMel_2003b}.

However, this distribution is not favorable to drive the
instabilities (\cite{FlYas1994a}, \yearcite{Fleishman1994c}),
because the instability related to the gaussian angular
distribution is rather weak being only slightly above the
threshold. The efficiency of the instability is set up by relative
importance of negative derivative of the distribution function
over the momentum modulus and positive contribution related to
angular gradient, which can be quantitatively described by the
ratio of the angular gradient of the pitch-angle distribution to
the angular distribution function itself. Although the angular
derivative in figure \ref{distrib} is not small, the peak value
occurs at those angles where the angular distribution function is
large, the ratio of these values is, therefore, small everywhere,
figure \ref{distrib} bottom. Thus, the gaussian pitch-angle
distribution can be stable against instabilities for rather broad
range of parameters.

Sin-N angular distribution \citep{Aschw1990} has very different
properties. While this function looks rather smooth (it is
continuous function together with its first derivative) and, thus,
attractive for the use in theoretical calculations, it cannot be
considered as a good approximation for the angular distribution
since this function has a very deep intrinsic problem. Indeed, if
we compare the behavior of the function and its first derivative,
we discover that the derivative decreases (with the decrease of
the pitch-angle) much more slowly than the function itself.
Particularly, the arrows in figure \ref{distrib} indicate the
region where the derivative is of order of unity, while the
function is much less than unity. Note, that the ratio diverges as
the pitch-angle decreases. Thus, sin-N function describes the case
of highly unstable electron distribution, which gives rise to ECM
instability operating significantly above the respective threshold
for any parameters of this function. As a result, the instability
exists for almost any shape of the momentum distribution of fast
electrons, and its efficiency has extremely weak dependence on the
electron spectral index. Such highly unstable distribution
function can hardly be built up under reasonable assumptions about
fast electron acceleration and transport.

This artificial property has already produced a few unexpected and
poorly understood results. In particular, the detailed numerical
calculations of the ECM quasilinear saturation \citep{Aschw1990}
did not ensure the entire maser saturation even after 50 growth
times: the peak value of the growth rate was reduced by less than
one order of magnitude for this time. The reason for this
surprising behavior is rather slow angular diffusion in the range
of small pitch-angles, where the distribution function is small
but the angular gradient is sufficiently large to provide the
instability and then support it for quite a long time. The
quasilinear relaxation occurs differently if the instability
operates closer to its threshold \citep{FArzner2000}.

Exponential loss-cone function (\ref{distr_mu_exp}) is free from
such kind of problem. From the theoretical point of view it has
only one disadvantage -- discontinuity of its first derivative at
$\theta=\theta_c$. Alternatively, a similar loss-cone distribution
but with faster, gaussian, decrease of the number of electrons in
the loss-cone (\ref{distr_mu_gauss}) is also attractive from the
theoretical point of view, since it is a function continuous
together with its first derivative.

\section{Incoherent Gyrosynchrotron Emission}

\begin{figure}
  \begin{center}
    \FigureFile(80mm,80mm){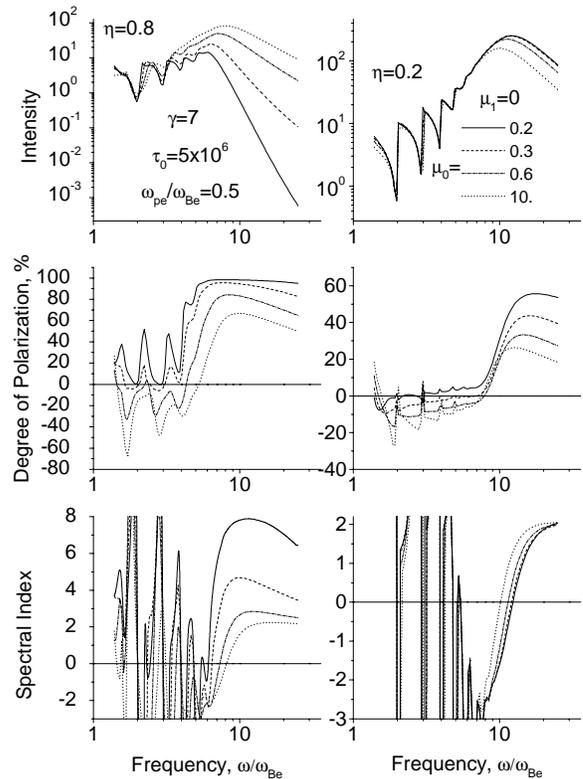}
  \end{center}
  \caption{GS radiation intensity, degree of polarization, and spectral index
vs. frequency for various $\mu_0$ in the Gaussian pitch-angle
distribution $f_2 \propto \exp\left[-(\mu-\mu_1)^2/\mu_0^2\right]$
of the loss-cone type, $\mu_1=0$}\label{gyro_1}
\end{figure}

The study of the GS emission produced by anisotropic pitch-angle
distributions requires exact calculation of the corresponding
emission and absorption coefficients \citep{FlMel_2003b}. This is
especially important for the directions along which a relatively
small amount of particles is moving, so adding up the
contributions from numerous electrons moving at unfavorable
directions should be done correctly. Figure \ref{gyro_1} presents
an example of intensity, degree of polarization and spectral index
of GS radiation produced by a gaussian angular distribution vs
frequency as the degree of the anisotropy (angular gradient)
changes \citep{FlMel_2003b}.

The effect of the anisotropy on the properties of GS emission
depends evidently on the viewing angle between the directions of
wave propagation and the magnetic field. For the quasitransverse
(QT) case, $\eta=0.2$ ($\vartheta=78\arcdeg$), right column, the
effect of the pitch-angle anisotropy on the spectral index (bottom
panel) is rather weak, while the intensity and degree of
polarization increase with anisotropy increases. However, the
anisotropy has exceedingly large effect on GS radiation produced
at a quasiparralel (QP) direction, $\eta=0.8$
($\vartheta=37\arcdeg$), left column. Indeed, the radiation
intensity changes by orders of magnitude compared with the
isotropic case. Respectively, the difference in the spectral index
in the optically thin frequency range is up to a factor of four
(left panel, bottom). Remarkable change of polarization occurs as
well: the degree of polarization increases in the optically thin
region  and can approach $100\%$ for sufficiently high anisotropy.
Furthermore, the sense of polarization can correspond to X-mode in
the optically thick region contrary to the isotropic case.

\begin{figure}
  \begin{center}
    \FigureFile(80mm,80mm){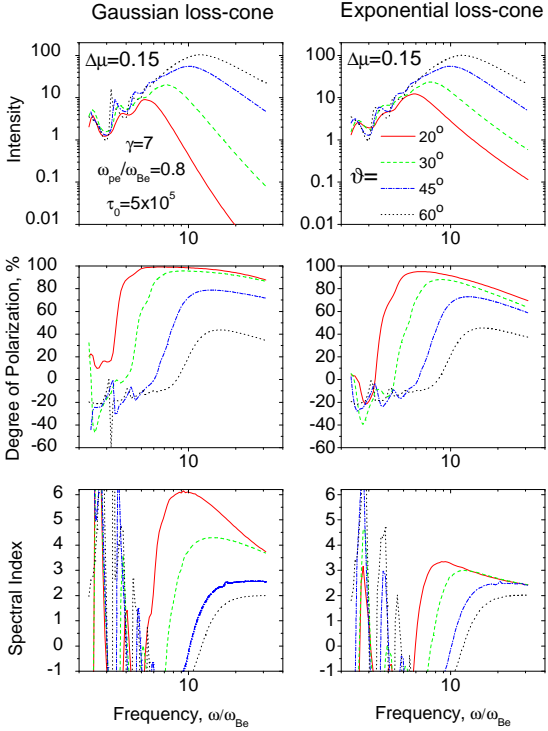}
  \end{center}
  \caption{GS
radiation produced by fast electrons with a power-law distribution
(with the index $\gamma=7$) over momentum modulus and gaussian
(\ref{distr_mu_gauss}) (left) or exponential (\ref{distr_mu_exp})
(right) loss-cone functions. The total intensity, polarization and
local spectral index are shown in the plot.}\label{gyro_2}
\end{figure}

Complementary, figure \ref{gyro_2} shows the intensity, degree of
polarization, and spectral index of GS emission observed at four
different viewing angles from the gaussian (\ref{distr_mu_gauss})
and exponential (\ref{distr_mu_exp}) loss-cone distributions (all
other conditions are the same). GS emission is evidently very
sensitive to the type of the angular distribution of fast
electrons. The difference between these two distributions are seen
most prominently in the spectral index plots: although the
spectral index can be as large as 6 for the gaussian loss-cone
distribution, it does not exceed 3 for the exponential loss-cone
distribution.

Thus, this high sensitivity of the results to the type of the
angular distribution can, in principle, be used to obtain
important constraints on the pitch-angle distributions present
during solar flares. We emphasize, that along with strong effect
on radiation intensity and spectral index, the anisotropy
influences the degree of polarization, which also might be applied
for quantitative diagnostics of the pitch-angle distribution.

Finally, we have to mention strong effect of pitch-angle
anisotropy on the low-frequency "harmonic" structure of GS
radiation \citep{FlMel_2003a}. However, detection of this
structure in the typical  non-homogeneous sources requires
high-resolution imaging spectroscopy, which is unavailable yet.

\section{Anisotropy Build up in the Magnetic Loops}

\begin{figure}
  \begin{center}
            \FigureFile(60mm,60mm){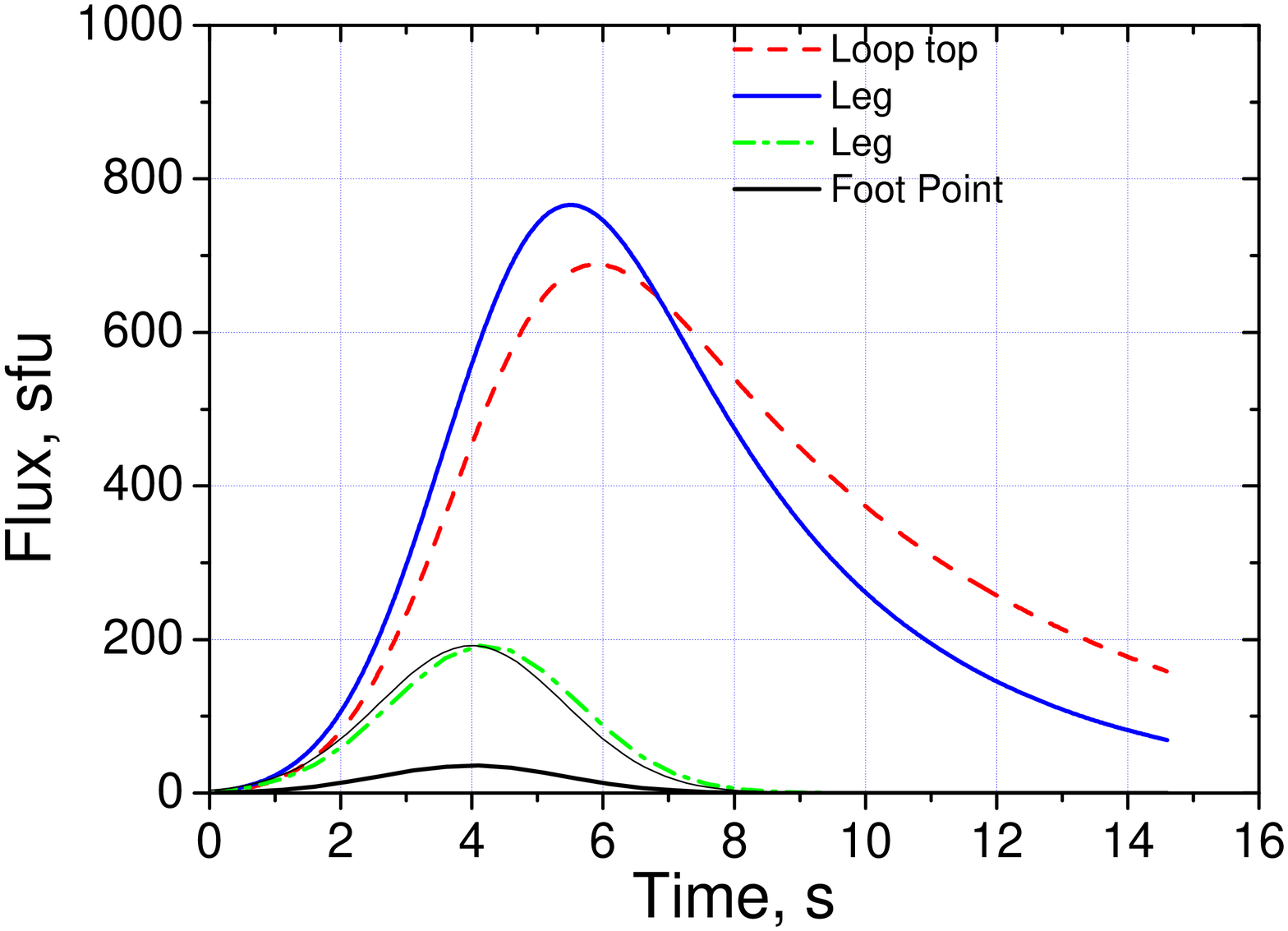}
            \FigureFile(60mm,60mm){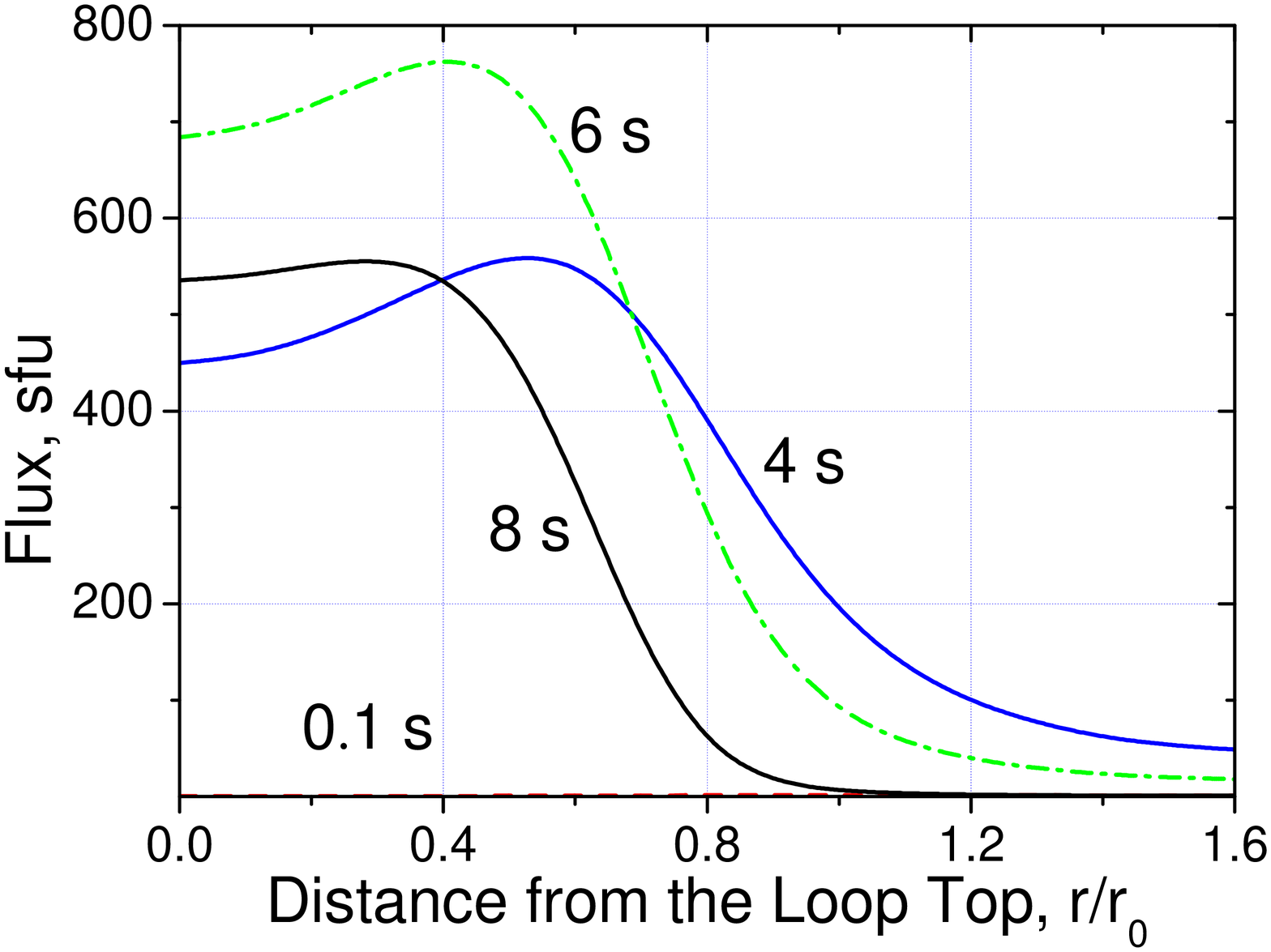}
            \FigureFile(60mm,60mm){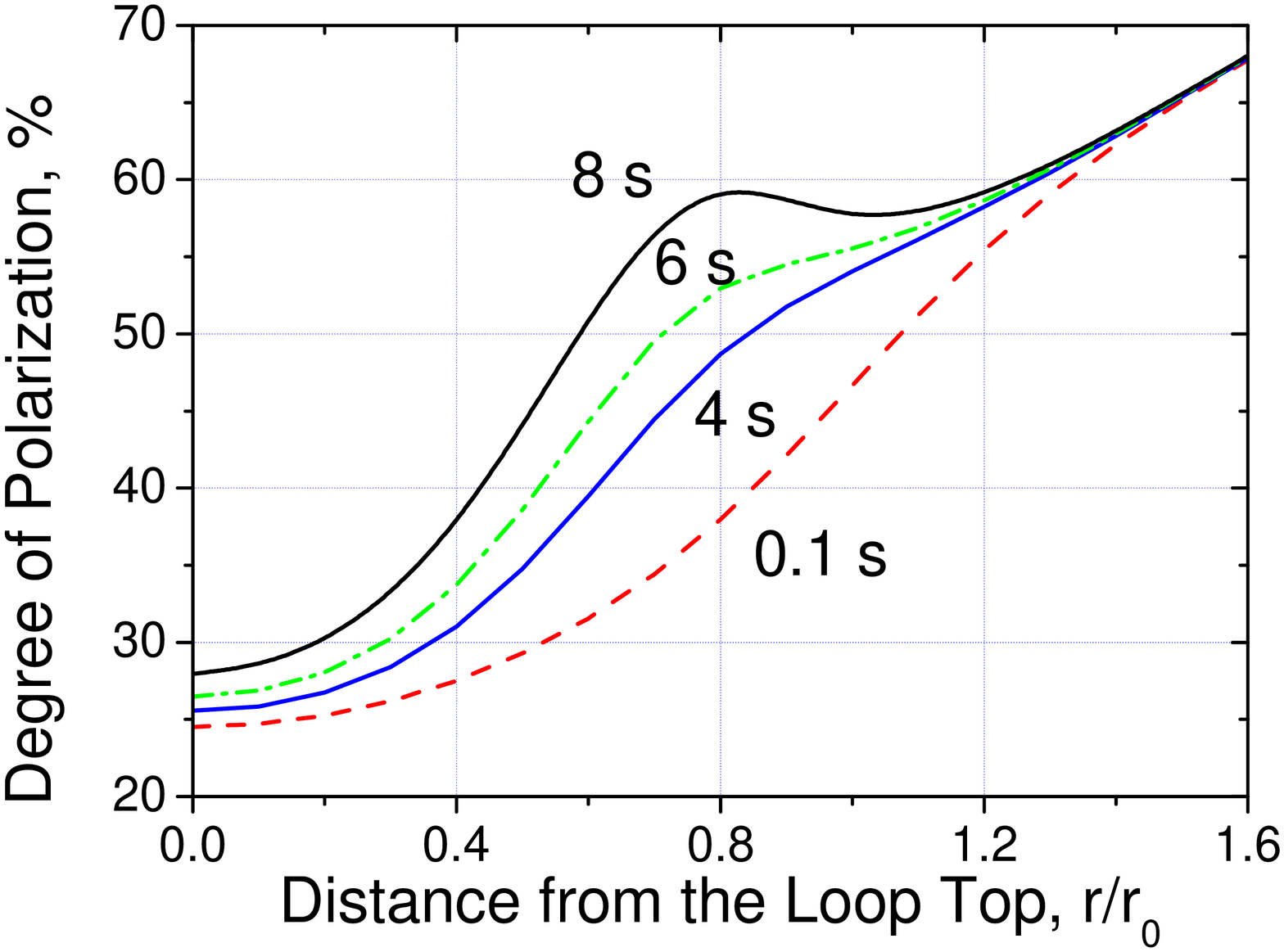}
  \end{center}
  \caption{Example of light
curves of the microwave emission from a few positions along the
loop  for the case of isotropic injection function and
angular-dependent life-time of fast electrons (top). The thin line
represents the injection profile. Corresponding  distributions (at
a few time frames) of radio brightness (middle) and degree of
polarization (bottom) along the magnetic trap. }\label{transp}
\end{figure}

The importance of transport effects
\citep{Kennel_Petchek_1966,Melrose_Brown_1976} for generation of
nonthermal emissions during solar flares has been recognized long
ago, in particular, the magnetic mirroring was argued to be a key
process to account for the loop-top brightness peak of the
optically thin GS emission \citep{Melnikovetal2002a}. This finding
implies that the anisotropy can actually be built up due to
transport effects even if the injection of fast electrons occurs
isotropically (although this does not deny the possibility of
anisotropic injection of fast electrons).

Typical treatment of the electron trapping at the magnetic loops
frequently explores  the idea of empty loss-cone due to particle
precipitation and losses at the dense foot-points in the weak
diffusion limit. Although this idea brings an apparent advantage
of the simplicity, it has also an important disadvantage of the
discontinuity of the distribution function over the pitch-angle
resulting in steep angular gradient, which affects strongly the
calculation of the absorption coefficients.

To avoid the problems with the discontinuity and account for
magnetic mirroring, energy losses and weak angular diffusion, we
put forward a new model solution of the transport equation:
\begin{equation}
\label{transport}
 f(E,\mu,s,t)=\int_{-\infty}^t \exp\left(
 -\frac{t-t'}{\tau(E,\mu,s)}\right) g(E,\mu,s,t')dt'
\end{equation}
where $g(E,\mu,s,t')$ and $f(E,\mu,s,t)$ are the injection and
distribution functions of the fast electrons at the radio source
vs energy $E$, the cosine of pitch angle $\mu$, position along the
loop $s$, and time $t$ . Here the phenomenological life-time
parameter $\tau$ can depend on the energy, position, and {\it
pitch-angle}.

For the weak diffusion regime we adopt that the life time $\tau$
display more or less fast decrease inside the loss-cone (when
$\mu$ approaches the unity). This ensures smooth behavior of the
distribution function, although the angular gradients can still be
large for a favorable combination of parameters, and the net
anisotropy can grow with time. Examples of the model radio
brightness distributions at the subsequent times as a radio burst
develops, and the spatially resolved light curves (at an optically
thin frequency) in the context of anisotropic transport are shown
in figure \ref{transp}. To reveal the net effect of the
angular-dependent transport, figure \ref{transp} assumes the life
time of the electrons to depend on the pitch-angle of the electron
only. It might be noted that the radio brightness peak moves
towards the loop-top as the burst develops in agreement with
observations \citep{Melnikovetal2002a}. Complementary, the
emission from different locations along the loop displays
systematic delay increasing from the foot-points to the loop-top,
which has also been widely observed with NoRH data
\citep{Melnikovetal2002b,Melnikov_2005,Bastian_2005}.

The dependence of the degree of polarization on time and its
spatial distribution are highly affected by the pitch-angle
anisotropy. Indeed, with isotropic injection, magnetic mirroring,
and weak angular diffusion adopted in figure \ref{transp}, the
anisotropy increases with time progressively, which gives rise to
the increase of the degree of polarization with time in the legs
and top of the loop. At the foot-points ($r/r_0 > 1.4$), however,
the angular distribution repeats the injected distribution since
the life time of the electrons at the foot-points is too small to
build up noticeable pitch-angle anisotropy.

\section{Observational Evidence of Anisotropic Pitch-Angle
Distribution in Flares}

\begin{figure}
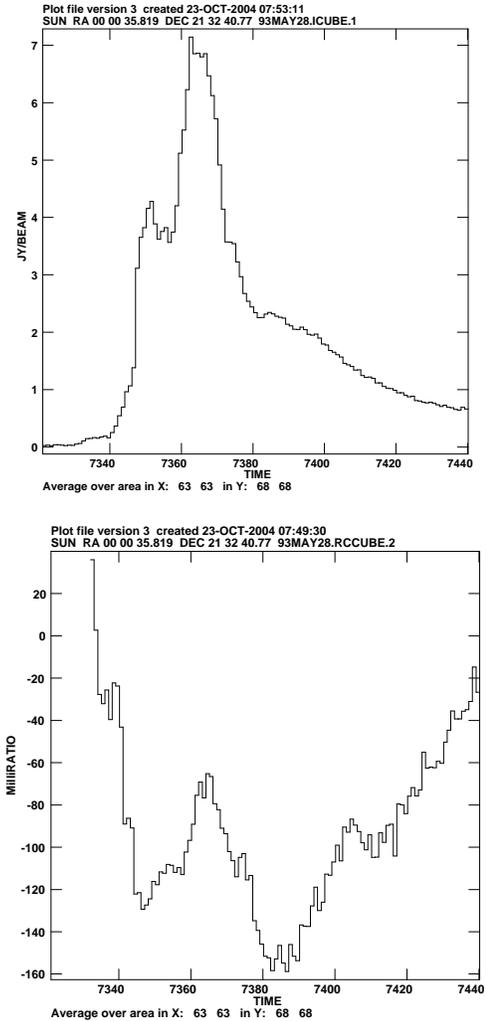

  \begin{center}
            \FigureFile(63mm,70mm){I3.ps}
            \FigureFile(66mm,70mm){RC3.ps}
  \end{center}
  \caption{A localized light curve (top) and and degree of polarization (bottom) for the 28-May-1993 event
  as observed at 17 GHz with NoRH (courtesy by T.S.Bastian). }\label{Deg_Pol}
\end{figure}

Besides already discussed spatial distributions of the radio
brightness and spatially localized light curves, behavior of the
degree of polarization typical for anisotropic pitch-angle
distributions is frequently observed with NoRH. An example of the
localized light curve (from one pixel of the magnetic loop leg)
and corresponding time profile of the degree of polarization for
the flare observed on 28-May-1993 is shown in figure
\ref{Deg_Pol}. Remarkably,  the absolute value of the degree of
polarization keeps growing during the main peak of the burst,
which is indicative for the increase of the anisotropy probably
due to particle losses through the loss-cone, since other
parameters affecting the polarization (magnetic field and the
viewing angle) cannot change much at the localized region of the
loop. The increase of the degree of polarization is followed by
its decrease at a later stage of the burst. Interestingly, this
turning point in the degree of polarization plot occurs at the
same time as the break in the light curve where rapid decay of the
radiation intensity slows down noticeably. This indicates  that
the losses due to precipitation into the loss-cone (which provides
the anisotropisation of the electron distribution) are giving way
to another decay process, probably, to Coulomb losses of already
accumulated fast electrons. The Coulomb losses result it harder
electron spectra, which generate GS radiation with weaker
polarization.

\begin{figure}
  \begin{center}
            \FigureFile(80mm,80mm){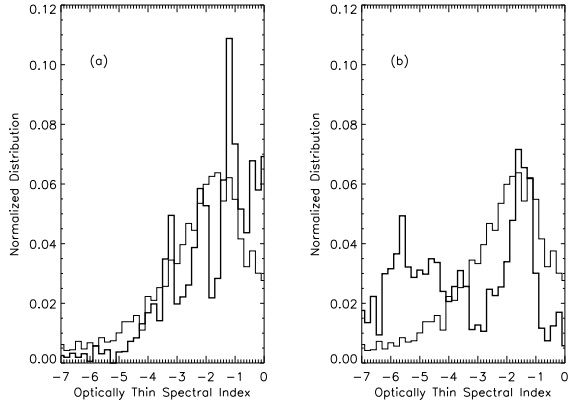}
            \end{center}
\caption{The histograms of the high-frequency microwave spectral
indices \citep{spikes} for all bursts recorded by OVSA during 2001
(thin line), superimposed on the corresponding histogram for
spike-producing microwave bursts observed at QT viewing angle
(left) and QP viewing angle (right), thick lines. Strong excess of
the steep spectra is observed in the QP case. } \label{thin_histo}
\end{figure}

A prominent spectral signature of the pitch-angle anisotropy has
been found from the detailed analysis of two-component radio
bursts composed of a high-frequency GS component and
lower-frequency coherent component \citep{spikes} observed with
OVSA. Indeed, the microwave sources (accompanying the coherent dm
spiky emission) observed at QP viewing angles to the magnetic
field, figure \ref{thin_histo}, displays remarkably softer spectra
at the optically thin region, than all other events on average,
which can only be understood within the effect of loss-cone
angular anisotropy on the GS spectrum. Moreover, all of the
microwave bursts were highly polarized in the optically thin range
\citep{spikes}, which is most probably related to the pitch-angle
anisotropy of the loss-cone type (see figures \ref{gyro_1},
\ref{gyro_2}).

The difference in the spectral indices for these two classes of
spike-producing microwave bursts (QT and QP) is as high as 3-4,
which implies some important constraints on the shape of the
angular distribution function. For example, the exponential
decrease of the number of electrons in the loss-cone
(\ref{distr_mu_exp}) turns  to be too slow to provide the observed
difference of 3-4 in spectral indices of optically thin GS
emission at QP and QT directions (figure \ref{gyro_2}, right
column). However, the gaussian loss-cone distribution
(\ref{distr_mu_gauss}) can evidently provide the required
difference in the optically thin spectral indices (figure
\ref{gyro_2}, left column). One can also note, that the gaussian
loss-cone distribution provides stronger degree of polarization in
the optically thin range than the exponential loss-cone
distribution.

\section{Electron Cyclotron Maser Emission}

It is clear and has been well-known for a long time, that the
pitch-angle anisotropy can drive instabilities, in particular, ECM
instability. In essence, ECM emission is a coherent GS emission
arising around lowest harmonics of the gyrofrequency when the
absorption coefficient changes the sign and turns to be the
amplification coefficient.

The theory of ECM emission in the solar corona has been
extensively developed for the recent years (see
\citet{Fleishman2003} for more detailed discussion and
references). Big efforts have been made to consider the linear
stage of the instability to specify the fastest growing wave-mode
as a function of the plasma parameter $
Y=\omega_{pe}/\omega_{Be}$, where $\omega_{pe}$ is the electron
plasma frequency and $\omega_{Be}$ is the electron gyrofrequency.

Then, a particular attention has been paid to the role of
absorption by the background coronal plasma in either emission
source or along the ray path (in the harmonic gyrolayers)
\citep{Sharmaetal1982,SharVla1984}. The absorption in gyrolayers
was found to represent a severe problem for the ECM emission to
escape the corona \citep{SharVla1984}, since the optical depth of
the second gyrolayer calculated with the maxwellian background
distribution is typically much larger than unity. However,
estimates of nonlinear absorption (when the background
distribution is modified by the incident powerful radiation)
\citep{Huang1987} and particle-in-cell simulations
\citep{McKeen_etal1990} show that even fundamental ECM emission
has a chance to escape. Nevertheless, it remains unclear if the
nonlinear absorption actually happens in solar conditions or not
(and, accordingly, which harmonics can really escape the source to
the observer).

Nonlinear stage of the ECM instability has also been studied. {\it
One-dimensional} particle-in-cell simulations
\citep{Pritchett1986} were found to agree well with a simplified
analytical approach \citep{Wuetal1981}. The one-dimensional
quasilinear saturation of the instability gives rise to a plateau
formation in the distribution function of fast electrons over the
transverse (to the magnetic field) velocity, providing, therefore,
an absolute stabilization of the electron distribution. This
means, that only the wave-mode with the largest growth rate can be
amplified substantially, while weaker growing wave-modes remain
practically unamplified.

The quasilinear relaxation is well known to occur rather
differently for the {\it two-dimensional} case
\citep{Akhiezer1974} relevant to the ECM emission. Indeed, the
relaxation due to the dominant mode affects the electron
distribution in such a way that the new (modified) distribution is
stable against this particular wave-mode. However, this new state
is not necessarily stable against generation of other wave-modes,
i.e., it is not absolutely stable distribution. Thus, other
wave-modes can well continue to grow until they start to affect
the electron distribution in their turn.

Two-dimensional fully relativistic numerical calculations of the
ECM quasilinear relaxation \citep{Aschw1990} revealed many
important properties of the relaxation process against the fastest
growing mode. However, the initial sin-N distribution function
adopted for the calculations has artificially large angular
gradient in the range of pitch-angles practically free from fast
electrons as has been already discussed in figure \ref{distrib}.
Quasilinear relaxation has further been studied numerically for a
particular set of the involved parameters when the lower-hybrid
wave-mode has the largest growth rate \citep{FArzner2000}. The
electron distribution was shown to remain clearly anisotropic
after the relaxation against the lower-hybrid waves. The new
anisotropic distribution is, however, stable against the
lower-hybrid waves, although unstable against the electromagnetic
wave-modes in agreement with the general results of the plasma
physics \citep{Akhiezer1974}. In addition, the time profile of the
generated wave energy density was found to resemble the observed
spike time profiles \citep{GuBenz1990}. Nevertheless, the ECM
saturation process has not been studied in detail yet, thus,
additional studies of the quasilinear relaxation against
electromagnetic wave-modes are necessary.

The role of coronal inhomogeneities both in ECM source and at the
ray path was analyzed.  Effect of MHD-waves on the longitudinal
transparency window \citep{Robinson1991a} and on the efficiency of
the mode conversion \citep{Robinson1991b} was studied. Then, ECM
line formation in a source with random inhomogeneities  was found
to produce either spectral broadening or splitting of the initial
peak depending on the strength of the inhomogeneity
\citep{Fleishman2004}. Mutual action of regular coronal
non-uniformity and random local inhomogeneities was considered for
a beam-like electron distributions \citep{Vlasov2002}.

In the previous section we concluded that the gaussian loss-cone
function (\ref{distr_mu_gauss}) is a good candidate for the
electron pitch-angle distribution in the main source of continuum
GS emission in the spike-producing events. Let us consider here
the main properties of ECM emission produced by fast electrons
with this angular distribution and power-law distribution over
momentum modulus to be compared with the observed properties of
the radio spikes appeared at the main phase of the burst. The
results presented below are obtained numerically from fully
relativistic calculations of the exact expressions of the GS
absorption coefficient.

\subsection{Basic parameters.} The growth rate value depends on many
parameters, therefore, we first selected a set of basic parameters
and then varied each of them successively. This approach allows to
discuss properly what effect is produced by each particular
parameter.

For our calculations we selected two basic values of the plasma
frequency to gyrofrequency ratio,
$Y=\omega_{pe}/\omega_{Be}=1,~1.3$. When $Y=1$, the most rapid
growth of the extraordinary waves occurs at the second harmonics,
$s=\omega/\omega_{Be} \approx 2$ and of the ordinary waves at the
fundamental, $s \approx 1$, while for $Y = 1.3$ the growth of O1
waves becomes ineffective, and it is giving way to O2 waves.

The angular distribution (\ref{distr_mu_gauss}) depends on two
parameters: loss-cone angle $\theta_c$ and the width of the
distribution $\Delta \mu$. We selected $\theta_c=60^o$ and $\Delta
\mu=0.15$. The later value allows the instability to operate in
the moderately-above-threshold regime. For the basic distribution
of fast electrons over momentum we selected the following
parameters typical for flares: $p_0/mc = 0.2$ ($E_{kin} \approx
10~keV$), $p_{br}/mc = 3$ ($E_{kin} \approx 1.1~MeV$) and
$\gamma=6$. It is assumed that at $p>p_{br}$ the power-law
distribution is giving way to exponential cut-off.

\begin{figure}
  \begin{center}
            \FigureFile(80mm,80mm){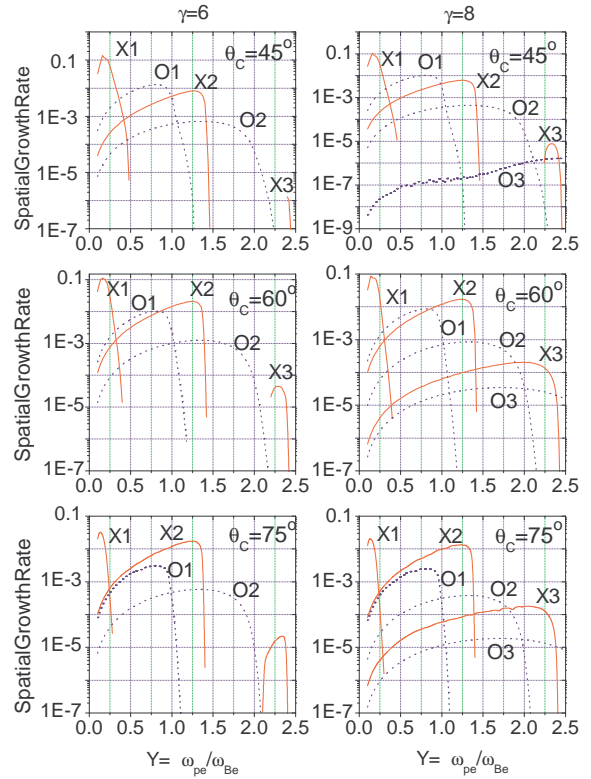}
            \end{center}
\caption{ Dimensionless spatial growth rates (normalized by ${\pi
\omega_{Be} \over 2c} {N_e \over N}$) for three first harmonics of
the gyrofrequency for various parameters: $\gamma=6$ (left
column), $\gamma=8$ (right column); $\theta_c=45^o$ (top),
$\theta_c=60^o$ (middle), $\theta_c=75^o$ (bottom). Set of basic
parameters is used: $\Delta \mu = 0.15$, $p_{0}/mc=0.2$,
$p_{br}/mc=3$.} \label{ECM_Y}
\end{figure}

\subsection{Effect of plasma density.} The effect of plasma density on the
growth rates at the lowest harmonics of the gyrofrequency was
repeatedly studied for various electron distributions
\citep{Sharmaetal1982,SharVla1984,Li1986,
WingDulk1986,Aschw1990,FlYas1994a,Fleishman1994c,
Ledenev1998,Stupp2000,Fleishman2003}. It has been established that
for small $Y < 0.2-0.4$ the fundamental extraordinary waves (X1)
have the largest growth rates. Then, for larger $Y$, the
amplification of X1 waves becomes inefficient and other waves
become dominant (O1, X2, ...). Our calculations  agree well with
this standard picture (since it is largely related to kinematic
restrictions).

The dependence of the growth rates on $Y$ (for the adopted
power-law momentum distribution and gaussian loss-cone angular
distribution) in the limits $0.1<Y<2.5$ is given in figure
\ref{ECM_Y}. The growth rate of each harmonics decreases rapidly
when the cut-off frequency of corresponding wave-mode exceeds the
respective multiple of gyrofrequency (i.e., when $Y>1$ for O1
waves, $Y>\sqrt{2}$ for X2 waves, $Y>2$ for O2 waves etc). The
absolute peak of the growth rate at a particular harmonics of the
extraordinary waves is noticeably larger than of ordinary waves at
the same harmonics. The absolute peak value decreases as the
harmonic number increases.

The range of $Y$ variation where a wave-mode is the fastest
growing mode depends on parameters of the fast electron
distribution. In particular, for small loss-cone angle value
(top), the growth rates of O1 waves are larger than of X2 waves
for $Y<1$. However, the difference between them decreases as
$\theta_c$ increases (middle) and X2 waves grow faster than O1
waves for large $\theta_c$ (bottom) in agreement with the finding
by \citet{WingDulk1986}. Respectively, O1 mode grows faster than
X2 mode if $Y<1$ for $\theta_c=45^o$, if $Y<0.85$ for
$\theta_c=60^o$, and never for $\theta_c=75^o$. Then, for
$Y>\sqrt{2}$ the O2 mode is the dominant, and for $Y>1.8$ the X3
mode may dominate (if it is unstable for a particular set of
parameters).

We note an interesting feature specific for our electron
distribution: the amplification of the third harmonics of both
extraordinary and ordinary waves is ineffective for hard electron
spectra (e.g., $\gamma=6$, left column) for most of the $Y$ range,
while it is effective for softer spectra (e.g., $\gamma=8$, right
column). The reason for the instability suppression for the hard
spectrum is the destructive contribution of (true absorption by)
high-energy electrons. For softer electron distributions the role
of high-energy electrons is less important allowing the
instability at the third harmonics. Note that for the gaussian
angular distribution used in (\cite{FlYas1994a},
\yearcite{Fleishman1994c}), the instability suppression by
high-energy tail of power-law distribution is important  even at
the fundamental and second harmonics, since the gaussian
distribution provides weaker angular gradients than considered
here gaussian loss-cone distribution (\ref{distr_mu_gauss}).

\subsection{Effect of angular distribution.}

\begin{figure}
  \begin{center}
            \FigureFile(80mm,80mm){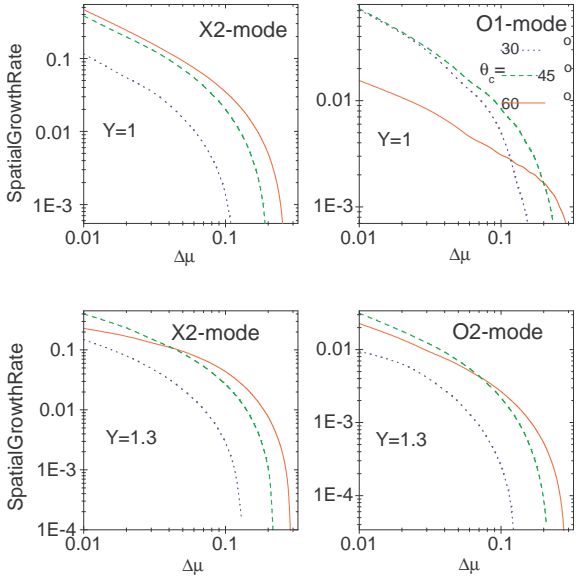}
            \end{center}
\caption{Spatial growth rates for X2-, O1-, and O2- modes vs width
of the loss-cone distribution $\Delta \mu$ for the basic set of
parameters and three $\theta_c$ values.} \label{ECM_mu}
\end{figure}

The dependences of the maximized (over the emission frequency and
angle) spatial growth rates on $\theta_c$ displays bell-shaped
curves for any mode. The most effective amplification of O1 waves
occurs around $\theta_c \approx 45^o$, while around $\theta_c
\approx 60^o$ for O2 waves and even higher values for X2 waves. No
qualitative difference is seen for various electron spectral
indices $\gamma$.

Of particular importance is the dependence on the width of the
distribution $\Delta \mu$, figure \ref{ECM_mu}. Evidently, all the
growth rates increase as $\Delta \mu$ decreases because this
corresponds to stronger angular gradients, and, thus, to stronger
instability. However, the growth rates decrease as $\Delta \mu$
increases first gradually and then rapidly. This rapid decrease
occurs when the instability approaches its threshold provided by
negative contribution of the momentum term ($\partial f/\partial p
=-\gamma f$) in the expression for the GS absorption coefficient.
When $\Delta \mu$ exceeds $0.1-0.3$ (depending on other
parameters) the instability disappears. We emphasize that this
happens when the pitch-angle distribution of fast electrons
remains clearly anisotropic. Thus, such anisotropic distributions
can generate standard GS emission without coherent ECM emission.

    The detailed shape of the curves depends on other parameters, like $Y$
and $\theta_c$. For X2 and O2 waves, large $\theta_c$ values are
preferable, while for O1 waves -- lower $\theta_c$ values.
However, for all considered cases, $\theta_c=60^o$ has the largest
threshold $\Delta \mu$ value.

\subsection{Effect of distribution over momentum.}

\begin{figure}
  \begin{center}
            \FigureFile(80mm,80mm){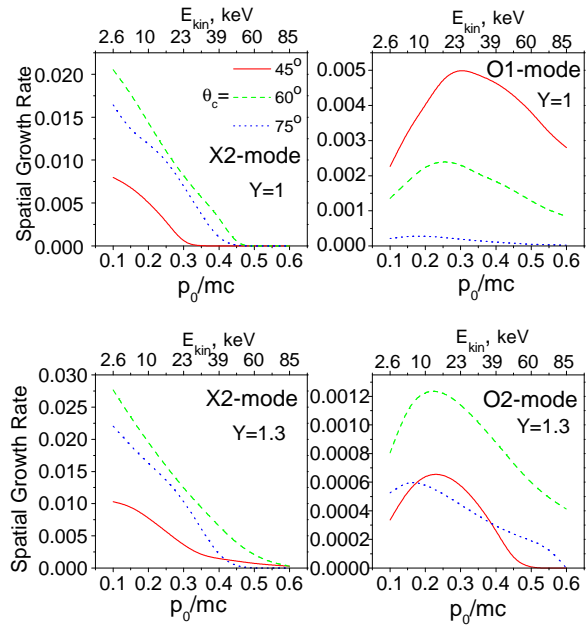}
            \end{center}
\caption{Spatial growth rates for X2-, O1-, and O2- modes vs
minimal momentum of fast electrons $p_0$ for the basic set of
parameters and three $\theta_c$ values.} \label{ECM_X0}
\end{figure}
The growth rates of all considered wave-modes display rather weak
dependence on $p_{br}$ for the selected set of basic parameters,
so we do not present the corresponding plots. We should note,
however, that this property of the ECM emission is model-dependent
and valid if the instability operates noticeably above the
threshold. If the instability is very close to its threshold, the
behavior of the momentum distribution in the high-energy range may
be important. In particular, this is the case of gaussian
pitch-angle distribution (\cite{FlYas1994a},
\yearcite{Fleishman1994c}).

The dependence of the growth rates on the minimal momentum $p_0$
is shown in figure \ref{ECM_X0}. Shape of the curves is clearly
different for various harmonics and wave-modes. X2 waves display
decreasing curves with increasing $p_0$, thus, low-energy
electrons are preferable to produce these waves. Wide loss-cones
($\theta_c=60^o,~75^o$) provide larger growth rates than narrower
loss-cone ($\theta_c=45^o$).

\begin{figure}
  \begin{center}
            \FigureFile(75.93mm,80mm){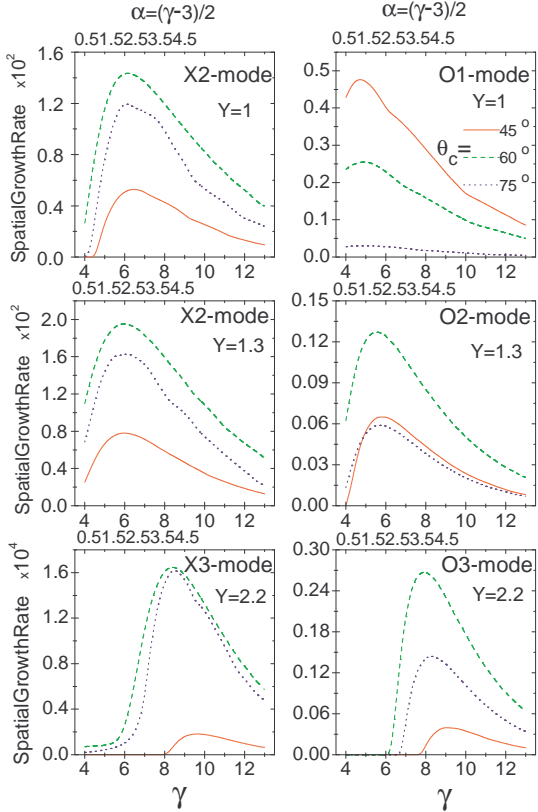}
            \end{center}
\caption{Spatial growth rates for X2-, X3-, O1-, O2-, and O3-
modes vs spectral index of fast electron distribution $\gamma$ for
the basic set of parameters and three $\theta_c$ values.}
\label{ECM_gam}
\end{figure}

Ordinary waves are generated most effectively if the low-energy
cut-off is about 20 $keV$ for O2-mode and about 10 $keV$ for O2
mode, although the corresponding peaks are rather broad. Narrow
loss-cone ($\theta_c=45^o$) is the most efficient to produce O1
waves and wider one ($\theta_c=60^o$) is preferable to produce O2
waves. The wide loss-cone of $\theta_c=75^o$ is less effective for
the both harmonics of the ordinary waves.

Considering together the plots in figure \ref{ECM_X0}, we can
conclude that different fractions of fast electrons are
responsible for generation of these different wave-modes. Thus,
quasilinear relaxation against one (the dominant) mode should not
necessarily provide saturation of all other wave-modes in
agreement with numerical quasilinear simulations
\citep{FArzner2000}.

The dependences of the growth rates on the electron spectral index
$\gamma$ are of particular importance, since they can be compared
with observations of accompanying continuum (microwave and HXR)
emissions. All the curves shown in figure \ref{ECM_gam} are
bell-shaped; peak of each curve corresponds to a favorable
$\gamma$ value specific for each wave-mode. The decrease of the
growth rates for harder spectra (smaller $\gamma$) is provided by
negative contribution of high-energy electrons at large harmonics.
The decrease for softer spectra (larger $\gamma$) is related to
the increase of the instability threshold that is proportional to
$\gamma$ for the power-law momentum distribution of the electrons.

The favorable $\gamma$ value depends on the harmonic number and
type of the wave. X2 waves have the peak around $\gamma=6$ (which
corresponds to the synchrotron spectrum with the index 1.5) for a
broad range of the involved parameters. A bit lower values are
favorable for O2 waves. However, O1 waves display the peak at
rather small values  $\gamma=4-5$.

It is remarkable that the favorable  $\gamma$ values for the third
harmonics are significantly larger than for the fundamental and
second harmonics: it is about 8 (synchrotron spectral index 2.5)
for O3 waves and even larger for X3 waves. Such a different
behavior of the curves for the different waves can in principle be
used to specify the contribution of various harmonics to the spike
radiation.

\section{Discussion}

We conclude the pitch-angle anisotropy to affect significantly the
radio emission from solar flares. First of all, we should mention
the strong effect of the anisotropy on the intensity and spectral
index of the emission in the optically thin region, which relates
primarily to the energy-dependent directivity (\ref{directi}) of
the synchrotron radiation by a single electron.

Currently, it is discovered with the NoRH data that microwave
emission from the foot-point parts of extended flaring loops
displays systematically softer spectra ($\Delta \alpha \sim$~
1-1.5 in the domain 17 GHz to 34 GHz) than from the loop-top part
\citep{Yokoyamaetal2002, Melnikov_2005}. Evidently, the
pitch-angle anisotropy can make an important contribution to the
observed spectral index variations. Indeed, if we take into
account the convergence of the magnetic field lines towards the
foot-points, we should expect a more anisotropic electron
distribution (due to a larger loss-cone) in the foot-point source
than in the loop-top source. Furthermore, for the disc flares the
foot-point source is observed at a QP direction, while the
loop-top source is observed at a QT direction, which results in
{\it systematically} softer spectra of GS radiation from the foot
points compared with the loop-top. Thus, the pitch-angle
anisotropy probably plays a key role in the observed variations of
the spectral index of the microwave radiation along the loop.
Moreover, the pitch-angle anisotropy has strong effect on
polarization properties of GS emission, which is widely observed
with NoRH.

Coherent GS emission (ECM) is also highly relevant for
understanding the current radio observations. The available data
about the radio spikes are pretty consistent with the idea that
the source of spike cluster is a loop filled by fast electrons and
relatively tenuous background plasma \citep{spikes}. The trapped
fast electrons (with a power-law energy spectrum and a loss-cone
angular distribution) produce continuum GS emission through the
{\it entire} loop, while each single spike is generated in a {\it
local} source inside this loop by ECM mechanism when some
favorable conditions are fulfilled. Most probably, the local
source is formed when the local anisotropy is increased compared
with the averaged one to produce ECM emission \citep{FlMel1998}.
The assumed fluctuations of the pitch-angle distribution of fast
electrons can be produced by the magnetic turbulence
\citep{FlMel1998,BaKar2001}. Accordingly, if the averaged angular
distribution is stable against ECM generation, the number of
spikes in an event should increase as the intensity of the
magnetic turbulence increases (and other equal conditions).
Complementary, the closer the averaged angular distribution to the
instability threshold, the larger the amount of the local sources
(and, respectively, the number of spikes) might be formed by the
same magnetic turbulence.

The scale of a single spike source is estimated (indirectly) as
$\sim 200~km$ \citep{Benz1985}; a wave should experience many
e-folding amplifications over this length to produce observable
spike with large brightness temperature, so this requirement reads
$\kappa \gg 5 \cdot 10^{-8}cm^{-1}$, where $\kappa$ is the spatial
amplification coefficient. We conclude (from the numbers in figure
\ref{ECM_Y}) that the required amplification can easily be
provided for the fundamental and second harmonics of both X- and
O- modes. Moreover, the amplification of X3 mode is also possible
if fast particles are numerous enough, $N_e/N \gsim 10^{-2}$,
while the amplification of O3 mode is less probable.

The dependence of the growth rate on the fast electron
distribution is of primary importance since it can be compared
with the observations. Indeed, the decrease of the growth rates
with the  $\gamma$ increase corresponds to smaller efficiency of
the ECM emission for softer spectra of fast electrons. The
correlation of this type is actually observed when spectra of
microwave or HXR emission simultaneous to spike bursts are
analyzed: the harder the fast electron spectrum the stronger the
averaged spike flux \citep{spikes}. The range of microwave
spectral indices is typically $\alpha =1-4$. Thus, we can conclude
that most of the spikes are generated at a harmonics not larger
than second. ECM emission at the third harmonics (see Figure
\ref{ECM_gam}) is the most efficient if $\gamma \approx 8$
($\alpha \approx 2.5$), therefore, the softer microwave spectra
should be preferable for the third harmonics generation in the
range $\alpha < 2.5$, which has not been observed for spikes yet.

Then, the dependences on the angular gradient value ($\Delta \mu$)
display clearly that  the instability disappears (as $\Delta \mu$
increases) when the pitch-angle distribution still remains highly
anisotropic ($\Delta \mu = 0.2-0.3$). Accordingly, the overall
distribution of fast electrons accumulated at the loop may be
rather anisotropic for a long time providing noticeably softer GS
spectra (for a QP viewing angle) than the isotropic distribution,
which is observed indeed \citep{spikes}.

{\small The National Radio Astronomy Observatory is a facility of
the National Science Foundation operated under cooperative
agreement by Associated Universities, Inc. This work was supported
in part by NSF grant AST-0307670 and NASA grant NAG5-11875 to New
Jersey Institute of Technology, and by the RFBR grants
No.03-02-17218, 04-02-39029.}


\end{document}